# A fast co-simulation approach to vehicle/track interaction with finite element models of S&C

Demeng Fan[1,2], Michel Sebès[1], Emmanuel Bourgeois[1], Hugues Chollet[1] and Cédric Pozzolini[2]

[1] Université Gustave Eiffel, COSYS, IFSTTAR, F-77447 Marne-la-Vallée, France
[2] ESI Group, Paris, France
demeng.fan@univ-eiffel.fr

**Abstract.** Simulations of vehicle/track interaction (VTI) in switches and crossings (S&C) require taking into account the complexity of their geometry. The VTI can be handled via a co-simulation process between a finite element (FE) model of the track and a multibody system (MBS) software. The objective of this paper is to reduce the computing effort in the co-simulation process. In the proposed approach, the VTI problem is solved inside the MBS software to reduce the computational effort in the track model as well as the flow of input/output between both modules. The FE code is used to supply the matrices of stiffness, damping and mass at the beginning of the simulation. An explicit time scheme is used with mass scaling. A good agreement is found between both approaches with a reduction of the computing time by a factor of 10. This new approach allows the optimisation of the design of S&C in further studies.

**Keywords:** Vehicle-track interaction, Switch and Crossing, Finite element.

## 1      Introduction

Simulations of vehicle/track interaction (VTI) in switches and crossings (S&C) necessitate the consideration of the geometrical complexity. Moreover, it may also be necessary to model the contact or material non-linearity. The VTI problem is addressed using a co-simulation process and split into two sub-problems: the vehicle dynamics is solved using a multibody system (MBS) software, and the track dynamics is solved using a finite element (FE) model. This co-simulation approach is considered in this paper using the MBS code VOCO. It allows using an existing FE model without requiring any reduction of the number of degrees of freedom (dof) or further assumptions.

The feasibility of co-simulation between an external FE code and VOCO has been demonstrated in a curve in the case of S&C [1], but the process may be time consuming. Track models directly implemented in MBS with an explicit scheme have shown their efficiency in terms of computation times [2] but are still limited to simple geometries. A new method is proposed in order to combine the advantages of both approaches, enabling to address complex geometries with a reasonable computer time.

Following the benchmark on S&C [3], a comparison of track model formulations for simulation of dynamic vehicle-track interaction in switches and crossings has been achieved [4]. The objective of this paper is to reduce the computing effort in the co-simulation approach by the new proposed approach. Simulations are carried out with the Swedish 60E1-R760-1:15 turnout [4]. Reference results are obtained with the "standard" approach which is described in the following section.

## 2    Methodology

### 2.1    Standard approach

The VTI problem is seen as the coupling of two sub-problems: vehicle dynamics and track dynamics. The principle of the coupling, as shown in Figure 1, is an exchange of data between the MBS code and the FE code. The excitation of the track results from the train dynamics, expressed by the wheel-rail (W/R) contact force. As shown in the left link of Figure 1, the W/R contact force is taken from the MBS code as an input for the FE code to assess the track displacement. In the right link, the displacements of the track under the wheel are introduced as a feedback in the track geometry [5].

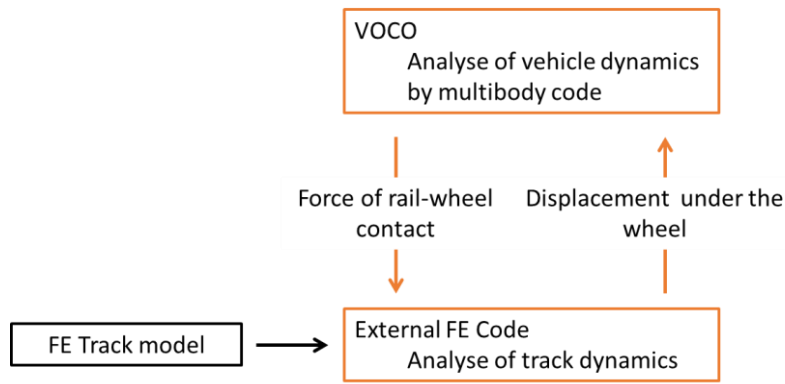

*Figure 1 Standard train-track interaction methods*

Wheel/rail contact is assessed with the semi-Hertzian method [6]. The tangent forces are evaluated by a variant of FastSim [7], which is based on the local properties of a strip instead of an ellipse. A functional approximation of this variant is used: this so-called Chopaya function may be also seen as an improvement of the Shen-Hedrick-Elkins method [8]. The wheel-rail contact force contains three forces and three moments, these latter ones being derived from forces with lever arms.

The standard approach is robust but still leads to long computation times. It seems reasonable to use independent time steps so as to perform FE analysis less frequently than the MBS analysis to reduce the calculation time. Attempts have been made without success to use a larger time step in the track module which is more computationally demanding than the vehicle block. Therefore, the current standard solution consists in using the same time steps for the MBS and the FE analysis to avoid numerical in-stabilities. The main bottleneck concerns the exchange rate of data in the time loop. On the one hand if the frequency of data exchange is too small, the coupling process may not converge. On the other hand, exchanging data too frequently implies longer simulation times. The MBS code VOCO uses an explicit time scheme which stable time step is smaller than the one required by the implicit FE code. An explicit scheme in the FE code would require such a small time step that it would block the coupling process due to the increase of input/output (I/O).

### 2.2    New approach

In the new method depicted in the flowchart of Figure 2, the third-party FE implicit solver is replaced by an explicit solver incorporated into the MBS code. The FE code is used at the beginning of the simulation to supply the matrices of stiffness, damping, mass, and boundary conditions. An interface is developed in which nodes without mass are eliminated in order to fulfil the prerequisite of explicit solvers. Due to the explicit time scheme, the stable time step proves to be smaller than the standard co-simulation. The new strategy consists in integrating a FE solver as a subroutine in the MBS code VOCO, hence suppressing the exchange of data through external files. The computational effort tends to decrease due to the limitation of I/O and the choice of an explicit time scheme.

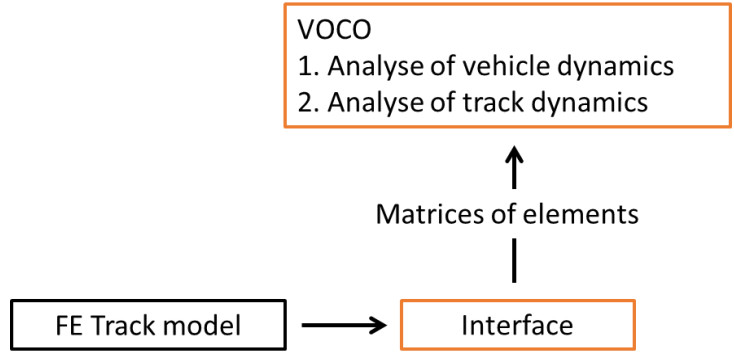

*Figure 2 New train-track interaction methods*

The track dynamics is solved via the equation of motion of the track model:

$$\boldsymbol{M}x + \boldsymbol{C}x + \boldsymbol{K}x = \boldsymbol{F}, \tag{1}$$

Where $\boldsymbol{M}$ is the mass matrix, $\boldsymbol{C}$ is the damping matrix, $\boldsymbol{K}$ is the stiffness matrix, $\boldsymbol{F}$ is the vector of W/R forces and $x$ are the track degrees of freedom. The time integration scheme is the Euler explicit scheme. The algorithm of the track module is:
a) Initialize displacements and velocities to zero.
b) Integrate velocities and displacements.
$$x_t = x_{t-\Delta t} + \Delta t x_{t-\Delta t},$$
$$x_t = x_{t-\Delta t} + \Delta t x_{t-\Delta t}.$$
c) Calculate the accelerations.
    Equation (1) may be rewritten as:

$$\boldsymbol{M}\boldsymbol{x} = \boldsymbol{F} - \boldsymbol{C}\boldsymbol{x} - \boldsymbol{K}\boldsymbol{x} \qquad (2)$$

The right side of equation (2) is assessed through a loop over the elements. Only the terms of the elementary stiffness and damping matrices associated to the active nodes are retained. So, the pointless contribution of the blocked dofs may be avoided. The W/R force $\boldsymbol{F}$ is applied to the closest nodes to the contact location, and then distributed by the cubic Hermite interpolation functions. The acceleration may be obtained from equation (2):

$$x_t = \boldsymbol{M}_l^{-1}(\boldsymbol{F}_{t-\Delta t} - \boldsymbol{C}x_{t-\Delta t} - \boldsymbol{K}x_{t-\Delta t}).$$

A lumped mass approximation $\boldsymbol{M}_l$ is adopted for the expression of the global mass matrix $\boldsymbol{M}$. According to the Courant–Friedrichs–Lewy (CFL) condition, the stable time step $\Delta t$ is inversely proportional to the maximal eigenfrequency and may be approximated from the ratio of the diagonal terms of $\boldsymbol{M}$ and $\boldsymbol{K}$ :

$$\Delta t = \frac{5}{\pi} \min \sqrt{\frac{M_i}{K_{ii}}}.$$

In the case of complex geometries such as S&Cs, the model may have small elements, significantly decreasing $\Delta t$. To solve this problem, the so-called 'mass scaling' technique is used where an extra mass is added to the nodes that have a significant influence on the stable time step. In order to control the level of approximation due to the mass scaling, the user may define the maximal mass $m_c$ to be added to each node. Nonetheless a compromise is required between the time step and the physics.

## 3    Case study

### 3.1    Swedish S&C with a linear track model

The studied cases are the same as in [4]. Both the through route and the diverging route are simulated. The FE track model is presented in Figure 3. The characteristics of the track model are reported in Table 1. Computing time with the standard approach is reported in Table 2. Simulations are carried out with a dual processor Intel Xeon Gold 5222 CPU @ 3.80GHz and a 128 GB RAM. The time step is 0.01 ms except for the diverging route in the switch panel where it is locally 0.005 ms.

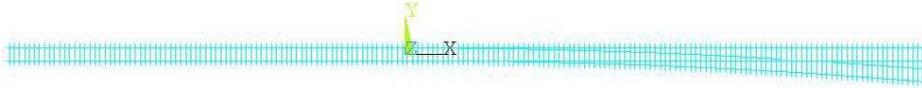

*Figure 3 FE track model of Swedish turnout in the international S&C Benchmark*

*Table 1 Characteristics of the FE track model*

| Definition | Value |
| --- | --- |
| Number of active dofs | 21740 |
| Total number of nodes | 16821 |
| Number of elements | 16225 |

### 3.2    Swedish S&C with voided sleepers

Voided sleepers are common defaults in ballasted tracks [5]. Modelling of voided sleepers introduce non-linearity in the track model. The same case study in a through route where 4 voided sleepers with a gap of 3 mm under the crossing nose are considered. This represents a case study in order to validate VTI in a nonlinear case. The maximum added mass per node $m_c$ is 10 kg. As shown in Table 2, introducing non-linearity in the standard approach results in more than doubling the calculation time.

*Table 2 Calculating time of standard co-simulation*

| Route | Linear track model | $\Delta t$(ms) | simulated time | $CPU_{ref}$ (h) |
|-------|--------------------|----------------|----------------|-----------------|
| Through | Yes | 0.01 | 1.85 | 17.56 |
| Diverging | Yes | 0.01/0.005 | 3.91 | 40.40 |
| Through | No | 0.01 | 1.85 | 43.65 |

## 4    Results

### 4.1    Swedish S&C with a linear track

The results of vertical W/R forces and vertical displacements under the wheel when $m_c$ is 50 kg for the through route are compared in Figure 4 and Figure 5, while $m_c$ is 10 kg for the diverging route in Figure 6 and Figure 7. The presented results focus on the transition zones in the switch and crossing panels. Indeed, the switch panel has two transition zones but only the second one at the loss of contact with the stock rail is shown. In terms of W/R forces (Figure 4 and Figure 6), a good agreement is found between both approaches, as it can be seen for instance by considering the P1/P2 peaks at the crossing panel. Only small differences in the peak values are visible in the switch rail forces at the loss of contact with the stock rail. In terms of rail displacements (Figure 5 and Figure 7) differences are more visible along the length of the track, which shows that the added mass has more influence on the rail displacement.

Table 3 presents the time step, the gain in computing time with respect to the standard approach, and the maximum difference with the standard approach in vertical stock rail displacements under the wheel, according to level of mass-scaling quantified by $m_c$ the maximum extra mass per node.

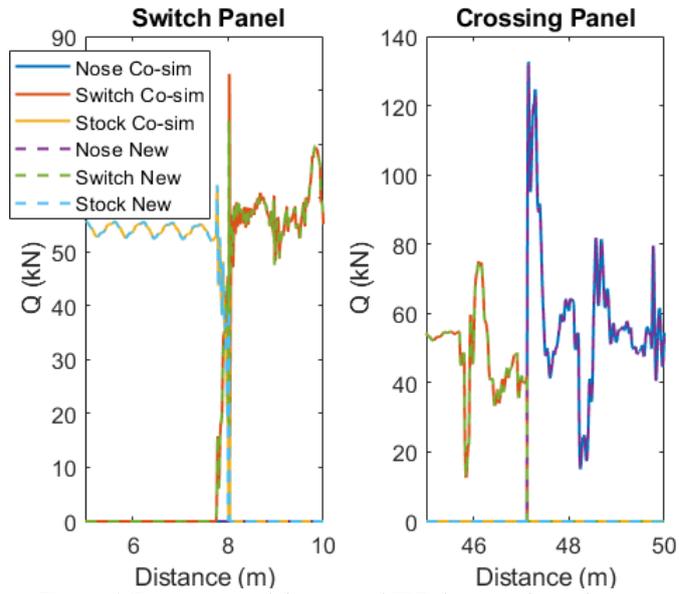

*Figure 4 Comparison of the vertical W/R force in through route*

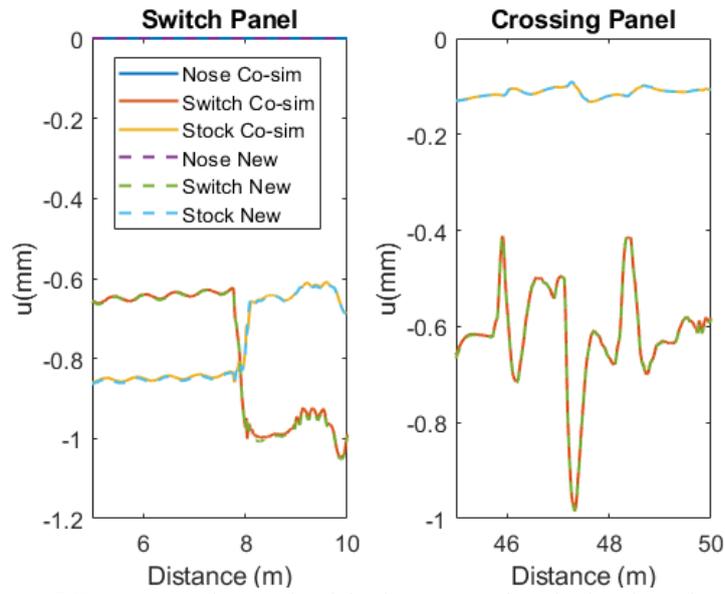

*Figure 5 Comparison of vertical rail displacement under wheel in through route*

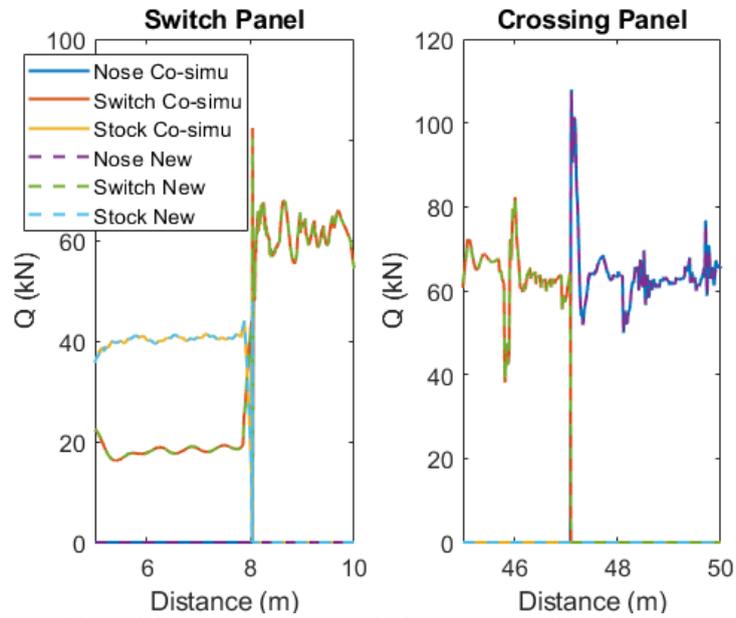

*Figure 6 Comparison of the vertical W/R force in diverging route*

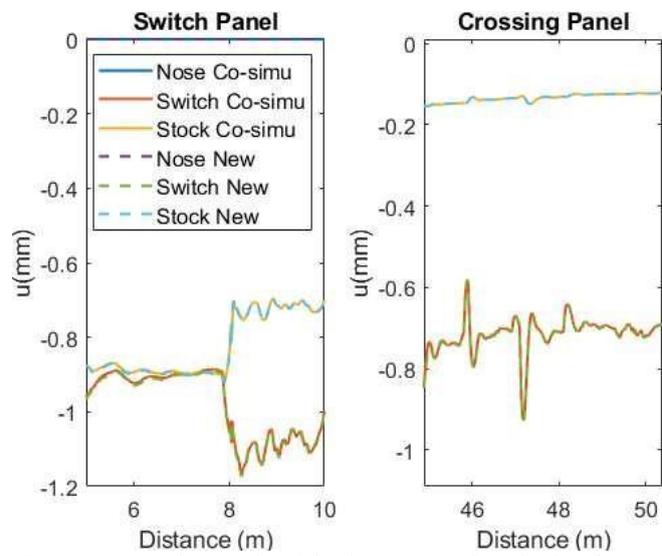

*Figure 7 Comparison of vertical rail displacement under wheel in diverging route*

*Table 3 Comparison in the through route according to the level of mass-scaling*

| $m_c$(kg) | $\Delta t$(ms) | $CPU_{ref}$ /CPU | max(u-$u_{ref}$)/$u_{ref}$ (%) |
|---|---|---|---|
| 20 | 0.0008 | 7.5 | 0.43 |
| 50 | 0.0014 | 11 | 0.61 |
| 100 | 0.0019 | 16 | 0.91 |
| 200 | 0.0028 | 23 | 1.53 |

Although the time step is significantly smaller to the one of the standard approach (see Table 2), the computing time is reduced by a factor of 10.

### 4.2    Swedish S&C with voided sleepers

The vertical force and vertical displacement in the crossing panel are compared in Figure 8 and Figure 9 respectively. The results agree well between the standard and the new approach. The effect of the voided sleepers is visible at the P2 peak of force (Figure 8) and in the magnitude of the rail displacements (Figure 9). A mass scaling threshold $m_c$ of 10 kg with a stable time step of 0.0006 ms has been used with the new approach. Still the gain of time is about 9 with respect to the standard approach. Because of the presence of elements with small length in the FE model, the stable time step without mass scaling would be 0.00003 ms. This shows the necessity of adopting a mass scaling strategy with the new approach.

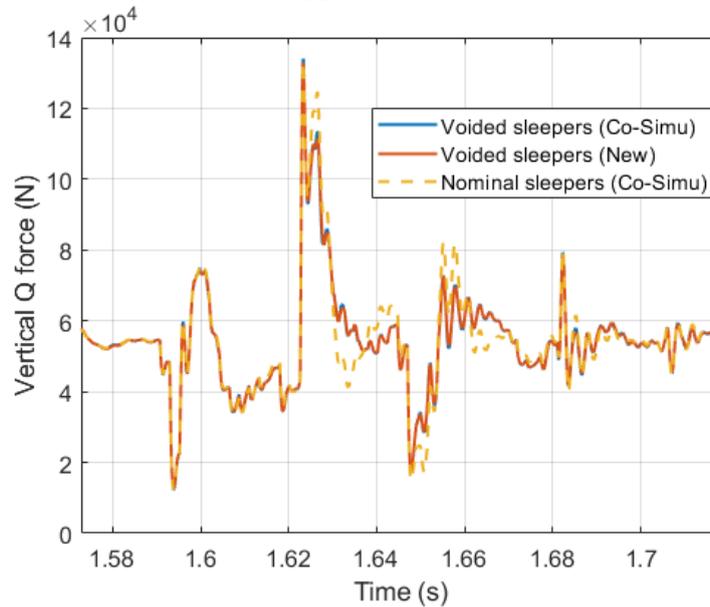

*Figure 8 Comparison of the vertical W/R force in through route on crossing*

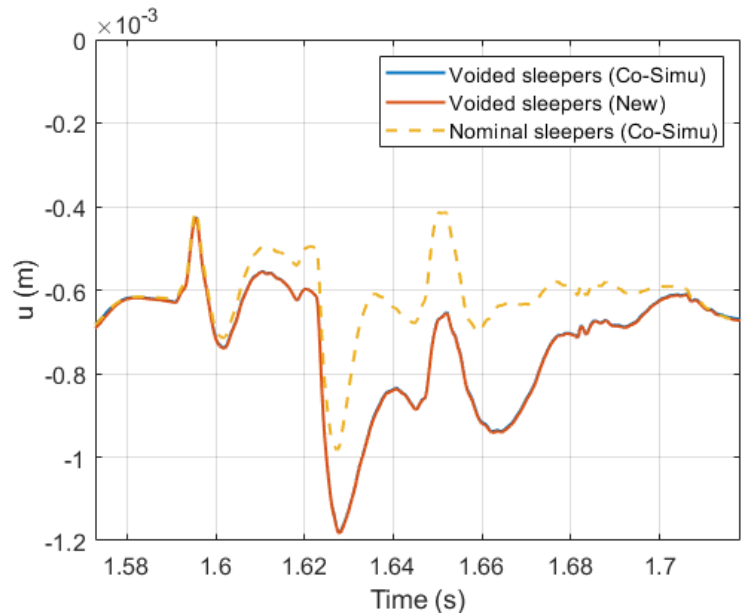

*Figure 9 Comparison of vertical rail displacement under wheel in through route on crossing*

## 5    Conclusion

A new method is proposed to address the VTI in turnouts within a reasonable computing time. The standard approach consists of a co-simulation with a third-party FE solver. Time saving is achieved thanks to an explicit scheme used in conjunction with a mass scaling technique, and the reduction of I/O. The external FE software is only used to provide matrices to the MBS. First test cases show a good agreement between both approaches with a reduction of the computing time by a factor of 10 both in the linear and non-linear cases. This new approach allows the optimisation of the S&C design in further studies.

### Acknowledgements

The work presented in this paper has been carried out with the generous support of ESI GROUP (Paris, France).

### References

1. M. Sebès, et al.: Adaptation of the semi-Hertzian method to wheel/rail contact in turnouts. Proceedings of the 24th Symposium of the International Association for Vehicle System Dynamics, IAVSD, Graz (Austria), (2015).
2. B. Eickhoff, L et al: Track loading limits and cross-acceptance of vehicle approvals. Proc IMechE Part F: J Rail and Rapid Transit. Vol. 229, No.6, pp. 710–728. (2015)
3. Y. Bezin, B.A. Pålsson, et al, Multibody simulation Benchmark for dynamic vehicle-track interaction in switches and crossings: Results and method statements Accepted to Vehicle System Dynamics (2021).

4. B.A. Pålsson et al. A comparison of track model formulations for simulation of dynamic vehicle-track interaction in switches and crossings (Submitted to VSD). (2020).

5. Y. Bezin, S. Iwnicki, M. Cavalletti, E. de Vries, F. Shahzad and G. Evans: An investigation of sleeper voids using a flexible track model integrated with railway multi-body dynamics. Proceedings of the Institution of Mechanical Engineers, Part F: Journal of Rail and Rapid Transit. vol. 223, pp. 597-607. (2009).

6. J-B.Ayasse and H. Chollet: Determination of the wheel rail contact patch in semi-Hertzian conditions. Vehicle System Dynamics. 43(3). pp.161–172. (2005).

7. Kalker, J. J: A fast algorithm for the Simplified Theory of rolling contact. Vehicle System Dynamics. Vo.11. pp. 1-13. (1982).

8. Z. Y. Shen, J. K. Hedrick, and J. A. Elkins: A comparison of alternative creep force models for rail vehicle dynamic analysis. The Dynamics of Vehicles on Roads and Tracks, Proc. of the 8th IAVSD Symposium. pp. 591-605. Cambridge (MA): Swets and Zeitlinger. Lisse. (1983).